\documentclass[manuscript,screen]{acmart}
\usepackage{graphicx}

\usepackage{caption}
\usepackage{subcaption}

\AtBeginDocument{%
  \providecommand\BibTeX{{%
    \normalfont B\kern-0.5em{\scshape i\kern-0.25em b}\kern-0.8em\TeX}}}

\setcopyright{acmcopyright}
\copyrightyear{2020}
\acmConference[fashionXrecsys '20]{fashionXrecsys '20: Workshop on Recommender Systems in Fashion, 14th ACM Conference on Recommender Systems}{September 22--26, 2020}{Virtual Event, Brazil}
\acmBooktitle{fashionXrecsys '20: Workshop on Recommender Systems in Fashion, 14th ACM Conference on Recommender Systems, September 22--26, 2020, Virtual Event, Brazil}
\begin{document}

\title[Outfit Generation and Recommendation]{Outfit Generation and Recommendation -- An Experimental Study}


\author{Marjan Celikik}
\email{marjan.celikik@zalando.de}
\author{Matthias Kirmse}
\email{matthias.kirmse@zalando.de}
\author{Timo Denk}
\email{timo.denk@zalando.de}
\author{Pierre Gagliardi}
\email{pierre.gagliardi@zalando.de}
\author{Sahar Mbarek}
\email{sahar.mbarek@zalando.de}
\author{Duy Pham}
\email{duy.pham@zalando.de}
\author{Ana Peleteiro Ramallo}
\email{ana.peleteiro.ramallo@zalando.de}
\affiliation{
\institution{Zalando SE}
  \city{Berlin}
  \country{Germany}
}

\renewcommand{\shortauthors}{Celikik, et al.}

\begin{abstract}

Over the past years, fashion-related challenges have gained a lot of attention in the research community. Outfit generation and recommendation, i.e., the composition of a set of items of different types (e.g., tops, bottom, shoes, accessories) that go well together, are among the most challenging ones. That is because items have to be both compatible amongst each other and also personalized to match the taste of the customer. Recently there has been a plethora of work targeted at tackling these problems by adopting various techniques and algorithms from the machine learning literature. However, to date, there is no extensive comparison of the performance of the different algorithms for outfit generation and recommendation. In this paper, we close this gap by providing a broad evaluation and comparison of various algorithms, including both personalized and non-personalized approaches, using online, real-world user data from one of Europe's largest fashion stores. We present the adaptations we made to some of those models to make them suitable for personalized outfit generation. Moreover, we provide insights for models that have not yet been evaluated on this task, specifically, GPT, BERT and Seq-to-Seq LSTM.

\end{abstract}

% https://docs.google.com/document/d/1b_pTrxSbyZlkW0VEVj4se7IYK6JYmQHew_x_8CIFUgg/edit?usp=sharing





\maketitle

%%
%% Custom styling and commands

% Bracket fix (see https://tex.stackexchange.com/questions/2607)
\let\originalleft\left
\let\originalright\right
\renewcommand{\left}{\mathopen{}\mathclose\bgroup\originalleft}
\renewcommand{\right}{\aftergroup\egroup\originalright}

%%
%% Content
\section{Introduction}

The role of fashion is constantly growing. In fact, over the last few years it has become one of the world's largest industries, with new trends, products, platforms, and brands constantly appearing. With the vast choice of items available in e-commerce, it has become increasingly difficult for customers to find relevant content, combine it, and match with a specific style.

Search and article recommendations are traditional systems that alleviate this problem. However, many consumers shop new items in order to complement an existing set of garments, or even a full outfit combination. Thus, these customers not only want to be recommended individual items, but a full outfit which is composed of a set of items of different types (e.g., tops, bottom, shoes, accessories), where these items have to be non redundant and visually compatible~\cite{DBLP:journals/corr/HanWJD17}. For this reason, over the past few years, many stylist-curated services have emerged, that help customers create outfits. However, these human-only-based approaches are not scalable in the growing fashion online market. Further, they may not leverage all the customer information and data that may be available.

Generating and recommending outfits is a huge challenge since it requires the items composing an outfit to be compatible with each other. There are multiple factors that define compatibility or fashion relationship such as brand, cut, color, visual appearance, material, length, and trends. Besides being compatible, the items should be personalized for the specific taste of each customer. 
Over the past years, a range of work has targeted these problems~\cite{FashionMeets2020}. Many researchers focused on pairwise compatibility~\cite{DBLP:journals/corr/VeitBK16, DBLP:journals/corr/McAuleyTSH15}, where the outfits are based on item-to-item compatibility. These approaches have the drawback that outfit compatibility is not computed on an outfit as a whole, but on pair-wise article combinations, which also makes them less suited for online serving due to high computation times. 

Recently, there has also been work inspired by ideas from the Natural Language Processing (NLP) community, by applying models such as Recurrent Neural Networks (RNNs)~\cite{DBLP:journals/corr/Lipton15} to generating full outfits~\cite{DBLP:journals/corr/HanWJD17}. This has the advantage that the outfit is considered as a whole and not only as pairs of items. However, considering an outfit as an ordered sequence poses unnecessary restrictions. More recently, a new stream of work has used the state-of-the-art model Transformer~\cite{DBLP:journals/corr/VaswaniSPUJGKP17} from NLP, in order to generate personalized outfits~\cite{POG2019}. The Transformer-based models BERT~\cite{DBLP:journals/corr/abs-1810-04805} and GPT~\cite{GPT2018} have not been tested on this task yet.

Even though there is a significant effort put into tackling the outfit generation and recommendation problem, to the best of our knowledge, there is no in-depth evaluation and comparison of the performance of different models on this task, including both personalized and non-personalized settings. Moreover, a lot of previous work provides results based only on open-source datasets~\cite{laenen2019attentionbased}, but not on real-world user data. In this paper we train and evaluate our models using datasets from Zalando\footnote{\url{https://zalando.com}}, one of the biggest online fashion retailers in Europe, with more than 500k articles and 32M active customers per year.

%This paragraph to be used when we provide the real life implementation for other occasion.
%With a few exceptions such as~\cite{POG2019}, previous work is not built and evaluated to be used in large-scale online setups. At Zalando, we provide a unique hub for digital fashion content in Europe, with tens of millions of active customers in Europe. We want to provide the best customer experience while generating outfits for our customers, which poses algorithmic challenges. However building a real-world outfit generation system poses several other challenges, which include operational and business constraints, cost, complexity and real-time, amongst others. Thus, when generating outfits, our systems not only need to maximize and improve our customer's experience, but also operational excellence. 

The contributions of our work can be summarized as follows:

\begin{itemize}
    \item We provide an in-depth evaluation and comparison of different algorithms on the outfit generation task using real-world user data. This includes both personalized and non-personalized approaches. The algorithms are Siamese Networks, Transformer, GPT, BERT, LSTM, and Seq-to-Seq LSTM;
    \item We adapt the language models BERT, GPT and Seq-to-Seq LSTM to \emph{personalized} outfit generation and extend the Siamese Nets architecture to outfit compatibility.
\end{itemize}

\section{Related Work}
\label{sect:related}

Fashion has become one of the world's largest industries. In fact, over the past few years it has gained a lot of attention both in the research community and the industry.  Wen-Huang Cheng et al.~\cite{FashionMeets2020} provide an overview of some of the main applications in the fashion domain, as well as a comprehensive survey of the state-of-the-art research.

% Outfit generation and recommendation is one of the biggest challenges, since not only the items in an outfit have to be compatible, but they also need to be personalized to match the customers' individual tastes. In this section we provide a summary of the various efforts available in the literature.

Plenty of previous work has focused on pairwise compatibility~\cite{DBLP:journals/corr/VeitBK16, DBLP:journals/corr/McAuleyTSH15}. To do so, many authors have used Siamese Networks~\cite{Koch2015SiameseNN}, which is a neural architecture that learns an item compatibility function, which in summary computes whether a set of items fit together or not. Veit et al.~\cite{DBLP:journals/corr/VeitKBMBB15} use them to learn style compatibility across categories, using data from Amazon. Vasileva et al.~\cite{DBLP:journals/corr/abs-1803-09196} propose an approach to learning an image embedding that respects an item's type, and jointly learns notions of item similarity and compatibility in an end-to-end model. McAuley et al.~\cite{DBLP:journals/corr/McAuleyTSH15} use a parameterized distance metric to learn relationships between co-purchased item pairs and used convolutional neural networks (CNNs) for feature extraction. More recently, Polonia et at.~\cite{DBLP:journals/corr/abs-1905-03703} leverage Siamese Networks for outfit compatibility, but opposed to previous work, the authors calculate the compatibility score using a fully-connected neural network. However, all these methods do not consider interactions among all the items in an outfit at once.

Inspired by the NLP community, several approaches have been applied to outfit generation. Kuhn et al. ~\cite{kuhn2019supporting} propose to use word2vec~\cite{mikolov2013distributed} to
learn a latent style embedding for each fashion item solely from the context in which an item appears, by exploiting the curations and expertise of their in-house styling experts. Lee et al. ~\cite{DBLP:journals/corr/abs-1708-04014} propose Style2Vec, a vector representation for fashion, which learns the representation of a fashion item using other items in matching outfits as context.

The use of RNNs has emerged as an alternative approach to item compatibility. Han et al.~\cite{DBLP:journals/corr/HanWJD17} use them to model outfit generation as a sequential process. However, considering an outfit as an ordered sequence poses unnecessary restrictions, since permuting the item positions should not alter their compatibility.

The Transformer~\cite{DBLP:journals/corr/VaswaniSPUJGKP17} is a sequence-to-sequence model which has been widely used in NLP. Based on this model, Chen et
al.~\cite{POG2019} present an industrial-scale Personalized Outfit Generation (POG) model that learns from the user-item and user-outfit interactions and generates a personalized outfit on the fly. Laenen and Moens~\cite{laenen2019attentionbased} propose an attention-based fusion method for outfit recommendation which fuses the information in the product image
and description to capture the most important, fine-grained product features. Other Transformer-based architectures such as BERT~\cite{DBLP:journals/corr/abs-1810-04805} or GPT~\cite{GPT2018} have also been used to tackle language-oriented tasks. However, there is no work that evaluates these two models on personalized outfit generation.

\section{Algorithms}
\label{sect:algorithms}

In this section we describe our outfit generation algorithms in detail. For clarity, we divide them into two groups: algorithms for \emph{item compatibility} (we will also refer to those as algorithms for non-personalized outfits) and algorithms for \emph{personalized outfits}. In the first group of algorithms, the learning problem is concerned only with fashion compatibility of a set of fashion items, while the second group of algorithms takes the user preferences into consideration, i.e., the item fashion compatibility is conditioned on the user.

Apart from briefly describing the original architecture that we based our models on, we include the changes we have implemented to some of them (e.g., Siamese Nets and GPT and BERT for item compatibility) in order to be able to adapt them to the outfit generation problem. % Furthermore, we discuss loss functions as well as inference design choices.%, namely, predicting articles as vectors in an article vector space or using full softmax layers to explicitly model a probability over a discrete set of items.

We define an \emph{outfit} $x=\{x_1, \dots, x_n\}$ to be a set of fashion items (garments and/or accessories) with compatible style, where each item can be related to any other item in the set. Depending on the algorithm, we define a user $u$ either as a sequence or a set of past actions (such as add-to-cart, add-to-wishlist, click, etc.), pertaining to items or by questionnaire answers (such as favorite brand, favorite colors, occasion, etc.).

\subsection{Item Compatibility}

In this section we describe various algorithms for the item compatibility problem, where the task is to learn which items are compatible and could fit together in an outfit. We start by generalizing the Siamese Nets~\cite{Koch2015SiameseNN} architecture, which we adapt to consider the compatibility of all items in the outfit rather than pairwise compatibility. Further, we describe adaptations of LSTM~\cite{LSTM}, BERT~\cite{DBLP:journals/corr/abs-1902-04094} and GPT~\cite{GPT2018}.

%\subsubsection{Style2Vec}

%TODO: still not decided whether should include https://arxiv.org/abs/1908.09493

\subsubsection{Siamese Nets}
\label{sect:siamese_nets}

The Siamese Nets architecture~\cite{NIPS1993_769} consists of two identical subnets with shared weights
that are inputs to a distance function used for compatibility matching. The distance function can
be fixed (e.g., Euclidean) or learned. The identical subnets serve as feature extractors that map
the input object into a latent encoding space that represents aspects of the input that are
important for compatibility matching. In case of learned similarity, these encodings are
concatenated and fed to the similarity block of the network to output the compatibility matching score.

We model the compatibility of two fashion items using a Siamese Nets architecture in a similar way.
The identical subnets do not share weights anymore since their respective inputs are different types of objects,
for example one for shoes and one for pants. We use sigmoid activation function and binary cross entropy to train the model, where a target
of 1 indicates that the items are compatible and 0 otherwise. As positive examples we use
stylist-created outfits. Negative examples are obtained by swapping uniformly at random up to $m$
items in a positive example with a random item, where $m$ is the length of the outfit.

Since using this approach models only the pairwise compatibility instead of the outfit as a whole, it has the
disadvantage that some items might not be compatible. To this end, we
generalize it by adding $n$ parallel subnets, one for each fashion category, for example, shoes, pants, dresses, and jackets. 
% For each subnet as well as for the similarity block we use fully-connected layers with ReLU activation.
We concatenate the outputs of each of the subnets as described above and include interactions
between them, namely the squared Euclidean distance and the Hadamard product
to obtain the vector $[x\: y\: (x-y)^2\: x\cdot y]$. The output of the network is computed in a similar way as before.

It should be noted that unlike the rest of the models described in this chapter that are based on
predicting score for all items in the vocabulary, Siamese Nets is a discriminative model and outputs score for a fixed set of items. Hence,
it involves computing forward passes on many candidate sets to find one with a high probability of
being an outfit. This has the drawback that the architecture is less efficient and it can pose difficulties in online settings where
outfit recommendations are generated in real time.

% TODO: consider moving the above paragraph to the experiments section.

\subsubsection{LSTM}
\label{section_lstm}

The work in \cite{DBLP:journals/corr/HanWJD17}~considers outfits as sequences instead of sets, where the order of fashion categories is fixed. The authors employ an LSTM~\cite{LSTM} to model item compatibility via learning the transitions between items as a proxy. Given a sequence of existing items, a forward LSTM is used to predict the next item in the sequence. Similarly, a backward LSTM is employed to model the previous item in the sequence in order to be able to construct a complete outfit. A zero item is appended to each sequence to serve as a stop token. Given an outfit $x$, the loss function is given by
\begin{equation}
\label{loss_lstm}
L\left(x; \Theta\right)=-\frac{1}{n}\sum_{t=1}^{n}\log\Pr\left(x_{t} \mid x_1, \dots, x_{t-1}; \Theta_f\right) -\frac{1}{n}\sum_{t=1}^{n}\log\Pr\left(x_{t} \mid x_n, \dots, x_{t+1}; \Theta_b\right)\,,
\end{equation}
where $\Theta=\begin{bmatrix}\Theta_f & \Theta_b\end{bmatrix}$ denotes the model parameters of the forward and backward model and $\Pr(\cdot)$ is the probability of seeing $x_t$ conditioned on the previous input. 

% We consider two different ways of estimating this probability. The first one is by employing \emph{full softmax} layer of all items in the training set. With this, the model is predicting the exact item. Since the softmax layer can become very large, we include only those items with frequency above a threshold $t$. In the second approach, we predict an embedding of the item and employ \emph{sampled softmax} on top of the output. Hence, $Pr(.)$ is computed by using a small sample of items $\chi$, for example, those in the batch of the next item:

% \begin{equation}
% \label{softmax}
% Pr(x_{t+1} | x_1, \dots, x_t; \Theta_f) = \frac{\exp\left(h_t x_{t+1}\right)}{\sum_{x\in \chi} \exp\left(h_t x\right) }
% \end{equation}

The outfit generation is autoregressive, i.e, the next item is predicted from an initial input set of items which then becomes the next input. To generate outfits with high probabilities, we employ \emph{beam search}, that works by maintaining a set of so-far most likely outfits based on perplexity defined as
\begin{equation}
\label{bert_perplexity}
PP(x;\Theta)=e^{L\left(x;\Theta\right)}\,.
\end{equation}

%In Section~\ref{s2s_lstm} we describe how to extend this algorithm by using sequence-to-sequence LSTMs to be able to condition the outfit generation on the customer preferences.

\subsubsection{Generative Pre-Training (GPT)}
\label{section_gpt}

GPT~\cite{GPT2018} is a popular autoregressive language model based on the Transformer architecture. It adopts the decoder part including the characteristic self-attention mechanism. 

Equivalent to the NLP use case, given an outfit $x$ we optimize the following loss function:

\begin{equation}
\label{loss_gpt}
L(x; \Theta)=\sum_i^{n} \log \Pr(x_i | x_{1}, \dots, x_{i-1}; \Theta)\,.
\end{equation}

where $\Pr(x_i \mid x_{1}, \dots, x_{i-1})$ is the conditional probability of an item $x_i$ given the previous items that is modeled using the Transformer decoder network with parameters $\Theta$.

The main difference to the original GPT language model is that items in outfits do not have an inherent order. Hence, we remove the positional encoding that is added to each token. The outfit sampling at inference time is done in an autogressive fashion similar to the LSTM.

% any other aoptions that might be interesting and are not obvious?
% * sorted vs non-sorted => more relevant for different tasks
% * having different input representations?

% Outfit sampling is done in an autogressive manner, first feeding the start token. The most probable , deriving the most probably item, feeding and then feeding each returned subsequence until a stop token is produced. 

%The outfit generation is autoregressive and works by generating the next item by feeding the initial one followed by the already predicted items until a zero token is generated.

% auto regressive property 

%to the language domain is that outfits do not . Therefore, we remove the positional encodings usually added to the input embeddings. Otherwise, we are equivalently modelling the probability of the next article ${x_i}$ in an outfit based on the previous $k$ articles. The network is trained on a set of example outfits maximizing the following likelihood:

% # FDNA
% An additional difference in the fashion domain is that the vocabulary is more volatile compared to language. Articles can for example go out of stock or disappear because of season changes. This can lead to problems with fixed vocabulary models if the tokens you predict are not valid anymore. One option to circumvent this is to predict the token embedding instead of the token probability itself. For this version of the model we calculate the loss according to~\cite{POG2019}, where we have a virtual softmax based on the true and four random negative sample embeddings: 

% {add equation from POG paper}

\subsubsection{Bidirectional Encoder Representations from Transformers (BERT)}
\label{section_bert}

BERT~\cite{DBLP:journals/corr/abs-1810-04805} is a masked language model based on the encoder part of the Transformer. It works by pre-training on unlabeled data using two tasks: fill-in-the-blank (FITB) and next sentence prediction. It has been shown empirically that BERT learns rich internal representations during the pre-training phase, which aid fast convergence and high accuracies on different downstream NLP tasks such as named entity recognition. In the following we outline the differences between the original BERT and our adaptation.

We modify the training objective and the output representation. Given an outfit, let $M=x_i$ be the event that item $x_i$ has been masked. The objective function of our BERT model for outfits can be written as
\begin{equation}
\label{loss_bert_for_outfits}
L\left(x; \Theta\right)=-\frac{1}{n}\sum_{i=1}^{n}\log\Pr\left(M=x_i \mid x\setminus\{x_i\};\Theta\right)\,,
\end{equation}
where $\Theta$ are the model parameters and $\Pr(\cdot)$ is the probability that the model assigns to $x_i$ being the masked item, conditioned on all other items in the outfit.

Similarly to GPT, we have removed the positional encoding of BERT. Furthermore, since there is no equivalent to the next sentence in the fashion domain, we remove the corresponding pre-training task altogether.

\subsection{Personalized Outfit Generation Algorithms}

In this section we describe algorithms for personalized outfit generation. This problem can be seen as an extension of the item compatibility problem which now includes \emph{context}. This context refers to any information external to the outfit. We distinguish between two main context types, namely \emph{customer actions} (e.g., clicks) and explicit customer preferences in a form of a \emph{questionnaire} (e.g., preferred brands, colors, prices, etc.)

To intuitively understand the advantage of providing context to the model, consider the following example. Suppose a customer has explicitly expressed that she likes casual footwear, such as sneakers. Also, she has previously clicked on items with colorful styles. If we can provide this context to a model it could infer that the customer prefers comfortable, colorful sneakers, and could generate a personalized outfit containing a pair of them.

In the remainder of the section, we start by introducing a generic approach to adapt any algorithm for the item compatibility problem to the personalized outfit generation problem. We then describe how the LSTM-based algorithm from Section~\ref{section_lstm} can be naturally extended to take the user context into consideration. Afterwards, we describe adaptations of the Transformer, and the BERT and GPT models for personalized outfit generation.

\subsubsection{Baseline Algorithm for Outfit Recommendation}

Any algorithm applicable to the item compatibility problem can be extended to personalized outfit generation in the following way. First, for each available item in the store, we compute $y$ outfits, where $y$ is sufficiently large to ensure these contain various styles and fashion attributes. Second, given a user $u$, we rank each outfit with respect to the user item browsing history, for example, by using learning-to-rank or simply by defining a similarity function between a user and an outfit. Such baselines approaches have been already considered in~\cite{conf/mm/HuYD15}. A particularly effective baseline based on nearest neighbours, defines the similarity between an outfit $x$ and user $u$ as follows
\begin{displaymath}
\label{eq:nearest_neighbors}
\frac{1}{|x|}\sum_{x' \in x} \max_{x'' \in U}\operatorname{sim}\left(x', x''\right)\,.
\end{displaymath}
where $\operatorname{sim}(\cdot)$ defines the similarity between two items, for example cosine similarity between item embeddings. The outfit with the highest score can be chosen as a personalized recommendation.

\subsubsection{Sequence-to-sequence LSTM}
\label{s2s_lstm}

Sequence-to-sequence LSTMs~\cite{NIPS2014_5346} map an input sequence of arbitrary length to an output sequence of an arbitrary length. This architecture is a straightforward application of the ordinary LSTM cell to general sequence-to-sequence problems. The first LSTM is used to read the input sequence to obtain a fixed-dimensional vector representation of the input. The second one is used to generate an output sequence, conditioned on the state of the first LSTM. 

In order to provide personalization, we provide the action sequence of a user $u$ as input to the first LSTM. The output is the outfit considered as a sequence, where the order of fashion categories has been fixed. Hence, the second LSTM learns an "outfit language model" conditioned on the user behavior. The loss and the outfit generation process is similar to that in Section~\ref{section_lstm} conditioned on $u$.

\subsubsection{Transformer}
\label{sect:transformer}

The Transformer~\cite{DBLP:journals/corr/VaswaniSPUJGKP17} is a powerful transducer model based on \emph{self-attention} with encoder-decoder structure that translates an input sequence to an output sequence.
In the Transformer adaptation of the personalized outfit generation problem, proposed in~\cite{POG2019}, the input to the encoder is the historical user behavior $u$ and the output of the decoder is an outfit $x$. Each item in the output is generated based on the previous items and the output of the encoder encoding $u$. Hence, the decoder learns the item compatibility conditioned on the encoded preference signal. The loss function of the Transformer is given by:
\begin{equation}
\label{loss_transformer}
L(x, u; \Theta)=-\frac{1}{n}\sum_{t=1}^{n}\log\Pr(x_{t+1} | x_1, \dots, x_t, u; \Theta)\,.
\end{equation}

It should be noted that apart from the user-item sequence that has limited length, the Transformer allows providing more global context, e.g., user segmentation or affinities for brands, styles, colors, etc. This can be done by assigning fixed positions in the encoder reserved for additional contextual embeddings. 

% TODO: introduce d_model (the internal dimensionality) so we can specify it as a hyperparameter in our experiments (Marjan: I think it's not necessary to mention it here as it's something well known about the transformer)

\subsubsection{Contextual BERT and GPT}
\label{sect:contextual_bert}
The BERT and GPT models for outfits described in Section~\ref{section_bert} and~\ref{section_gpt} are non-personalized, i.e., regardless of customer preferences or interactions, they generate the same outfits. However, we want to be able to \textit{condition} these models during inference in order to generate better suited outfits for each customer. The context we are using is information about the customer such as season, gender, age, weight, height, preferred brands, preferred colors, and other summarized customer information. We therefore extend both models and make them contextual in the following way.

We embed the context into a vector space which has the same dimensionality as the item embedding. For BERT this context is appended as an additional token and for GPT it is added as a start token. In both cases the models can attend to the context vector and utilize it for prediction. This method resembles the work in~\cite{DBLP:journals/corr/abs-1812-06705}, where binary information about the sentiment of a sentence is injected into BERT. 

% For both BERT and GPT the objective function changes to maximizing the probability of predicting the ground-truth item based on the rest of the items in the outfit and the newly introduced context.

% in the following way: in addition to having access to (some of) the other articles in the outfits the models are provided with the context. The objective is then to maximize the probability of predicting the ground-truth article based on other articles \textit{and} the newly introduced context.

While GPT can be naturally used to sample outfits autoregressively, BERT has originally not been designed for generative tasks~\cite{DBLP:journals/corr/abs-1810-04805}. Recent works such as~\cite{DBLP:journals/corr/abs-1902-04094}, however, suggests employing Gibbs sampling to retrieve full-length sentences from BERT. We adopt it by iteratively masking out positions in a randomly initialized outfit and using a trained BERT model to find replacements for them. 
% The context is present at all times so the outfit that is being constructed iteratively is conditioned on it. 
Note that many\footnote{The exact number of forward passes for sampling a single outfit from BERT depends on the implementation. We found it to be ideally at least an order of magnitude higher than the outfit length.} forward passes are required for this method, while GPT can generate an outfit of length $n$ in $n+1$ forward passes.

\section{Experiments}
\label{sect:experiments}

In this section we provide offline results and insights on the performance of different algorithms that were introduced in Section~\ref{sect:algorithms}. We first introduce the datasets and the features and then evaluate the non-personalized algorithms followed by the personalized ones.
% We start with the non-personalized algorithms, to then report on the performance of the personalized models. 

% TODO: add frequency thresholds for non-personalized and personalized algorits

\subsection{Datasets}
\label{sect:datasets}

In this section we present the datasets used to train and evaluate our models. They come from Zalando, a hub for digital fashion content in Europe. In the online shop, customers can purchase or seek for inspiration about garments and style via, for example, outfits.

The hand-crafted outfits we use to train our models are created in three different ways: (1) outfits from content creators (such as stylists) on the website (Shop the Look, STL), (2) influencer-created outfits (Get the Look, GTL)\footnote{\url{https://en.zalando.de/get-the-look-women}}, where influencers assemble their own outfits, and (3) via Zalando's personalized styling service Zalon\footnote{\url{https://zalon.de}} where stylists create outfits customized for each customer individually.

\begin{itemize}

\item \textbf{Zalando outfit dataset (GTL and STL)}: This dataset consists of around 250k hand-created outfits that have been published on Zalando, containing a total of 1M distinct items. This includes STL styled by Zalando creators, provided as an inspirational supplement to the item on the product detail page, and the influencer-created outfits available on GTL. Each of these outfits is composed of a single item per body part that can be worn together, occasionally accompanied by a fashion accessory.

\item \textbf{Zalon outfit dataset}: A dataset of around 380k recent outfits, each of which has been handcrafted for a specific customer by a professional stylist. A Zalon stylist assembles an outfit based on \emph{questionnaire} answers that a customer provides, where they express their style preferences, provide body features to the stylist, and specify price expectations. Based on this information, the stylist creates a personalized box consisting of two outfits, where each consists of up to seven articles, for example shoes, pants, t-shirt, sweater, and jacket. There can be multiple articles of one type, for example multiple pants or tops, but all of them are compatible amongst each other. Zalon's dataset contains about 30k distinct articles from the Zalando fashion store. To restrict ourselves to this limited set of distinct articles, we removed the long tail of articles which appear less than eight items.

\end{itemize}

In Table~\ref{table:outfitdataset} we show a summary of the two outfit datasets we just described. With the previous datasets, we can solve the compatibility problem. However, in order to be able to cater for the specific taste of our customers, we also make use of customer context. The following two datasets contain user specific data in the form of clicks and questionnaires:

\begin{table}
    \caption{Comparison of the key properties of Zalando's GTL \& STL and Zalon's outfit dataset.}
    \label{table:outfitdataset}
    \begin{tabular}{l l r r r}
        \toprule
        Dataset & Personalization & \#Outfits & \#Articles & Avg. outfit length \\
        \midrule
        Zalando GTL \& STL & Click history & 251,891 & 64,748 & 4.50 articles \\
        Zalon & Questionnaire & 380,808 & 30,619 & 4.96 articles \\
        \bottomrule
    \end{tabular}
\end{table}

\begin{itemize}

\item \textbf{Zalando click dataset}: This dataset consists of user click actions (clicks on articles, additions to the wishlist, etc.) on a single item and user actions on whole outfits that are available on Zalando. In total there are close to 1M outfits per year created on the Zalando web page available to approximately 32M active customers. We aggregate the past actions per user over a period of one month and create training samples consisting of outfit interactions together with the preceding item actions such as click and add-to-wishlist. The outfit actions are taken into consideration only if there are at least five item actions preceding it and contain at least four items, excluding accessories. We exclude fashion items that are rare and occur less than three times in the action data. This way, we obtain around 6M valid training samples that contain around 200k distinct outfits and 100k distinct items.

\item \textbf{Zalon questionnaire dataset}: In Zalon, each customer that requests a personalized outfit needs to fill a detailed questionnaire, which gives information about style preferences, provides body features to the stylist, and specifies price expectations. The total number of features we collect is over 30, with some examples of questionnaire fields being the shoe size, no-go dress types, favorite brands, favorite colors, hair color, body height and weight, and the occasion for which an outfit is needed. The total amount of questionnaires for training our models is around~250k.

\end{itemize}

Our datasets are proprietary and cannot be released for customer privacy reasons. The Zalon dataset is distinct from other datasets in the domain because of its rich questionnaire features which contain significantly more information about a customer than their click or purchase history. It is therefore especially promising to use in combination with personalized models.

\subsection{Item Representation}

For all our models, we use the same item representation that contains a 128-dimensional image embedding extracted from the penultimate fully-connected layer of a fine-tuned ResNet-50 CNN computed from the packshot item image. This vector is then concatenated with a vector of learned embeddings of categorical item attributes, in particular: \emph{category, brand, season, color, gender, material} and \emph{pattern}. We use a softmax layer to predict a probability distribution of a subset of the full vocabulary of items appearing in the training and test sets. We keep only the items with frequency larger than a predefined threshold of 8 occurrences.

\subsection{Non-Personalized Models}
\label{sect:experiments_item_compatibility}

In this section we present the results of our experiments on the non-personalized algorithms. We first describe the implementation details and define the metrics followed by evaluation on both the Zalando and the Zalon outfit datasets introduced in Section ~\ref{sect:datasets}

\subsubsection{Implementation Details}

% We represent each item by using embeddings extracted from the penultimate fully-connected layer of a fine-tuned ResNet CNN together with learned embeddings of textual item attributes such as brand, color, silhouette, pattern, occasion, season, etc.

%For all our models, we have used the same item representation: a concatenation of a 128-dimensional  image embedding extracted from the penultimate fully-connected layer of a fine-tuned ResNet CNN, as well as learnt embeddings of categorical item attributes, in particular: category, brand, season, color, gender, material and pattern. In all of our models we use soft-max layer to predict the entire vocabulary of items from the training and test sets.

% The additional features are amongst others an article's commodity group (jeans, shirt, ...), brand, season, color, gender, material, and pattern. They are embedded in a 77-dimensional article feature vector.

% Architecture
\begin{itemize}
\item \textbf{GPT} and \textbf{BERT}: We use four layers with eight attention heads each and set the model dimensionality $d_\text{model}=128$. We use batch size of 512 and a dropout rate of 1\%.
% Adam with
% The training runs for 125 epochs.

\item{\textbf{Siamese Nets}}: We use two fully-connected layers for the feature-extractor subnets and two fully-connected layers for the item compatibility part of the network. Each layer has 64 ReLU units. We generate the negative samples by randomly changing from one up to $n$ items in each outfit in the training set, where $n$ is the size of the outfit. We use batch size of 32.

\item{\textbf{LSTM}}: We use the setting from~\cite{DBLP:journals/corr/HanWJD17}: a single-layered LSTM cell with 512 hidden units and a dropout of 0.3. We train we batch size of 64.

% TODO: model parameter count comparison

\end{itemize}

We randomly split our data into 90\% train and 10\% validation data. 

% The train data is identical for all metrics, but we use different validations for the different metrics, which is described below.

\subsubsection{Metrics}

To evaluate the quality of the non-personalized models, we adopt three well-known metrics:

\paragraph{Perplexity (PP)} The perplexity is a common metric in the NLP domain. It reflects how well the model has learned an underlying distribution in an autoregressive fashion. In our case, a low perplexity indicates that the model is performing well at sequentially generating samples from the approximated outfit distribution. For a single outfit, the PP is defined based on the average cross-entropy (CE) as 
\begin{align}
\label{perplexity}
\operatorname{CE}\left(x;\Theta\right)&=-\frac{1}{n}\sum_{t=1}^{n}\log\Pr\left(x_{t} \mid x_1, ..., x_{t-1}; \Theta\right)\,,\\
\operatorname{PP}\left(x;\Theta\right)&=\exp\left(\operatorname{CE}\left(x;\Theta\right)\right)\,,
\end{align}
where $x$ is a ground truth outfit. To report the PP for the validation dataset we average it across the outfits. In an effort to make BERT comparable to GPT, we compute the perplexity for BERT by masking every item once and removing the respective context to its right.

\paragraph{Fill In The Blank (FITB)} The FITB recall at rank $r$, also abbreviated as FITB@$r$, measures the model's ability to complete an outfit where one item was masked out. It represents the probability that the ground-truth article is among the top $r$ predictions made by the model. In our case, we compute r@1, r@5, r@25, and r@250.

We implement FITB for GPT as follows. First, the masked item $x_i$ is removed from the outfit and the remaining items $x_{1}, \dots, x_{i-1}, x_{i+1}, \dots, x_{n}$ are fed into the network. The network's prediction at position $n$ is then interpreted as prediction for the masked-out item. Note that this is only reasonable if GPT is trained with randomly shuffled outfit sequences. To clarify this, assume that GPT is trained with outfits which are sorted as follows: shoes, pants and shirts. If the pants are removed and the model is presented with shoes and a shirt to predict the missing pants, this would constitute an out of distribution case since the model has never seen this combination in this order before. On the other hand, if the outfits are shuffled, the model has likely already seen such combinations.    

\paragraph{Compatibility Prediction (CP)} The outfit compatibility metric evaluates a model's capability at distinguishing compatible from non-compatible outfits. For each compatible outfit we generate a non-compatible example by replacing one item at a randomly selected position by another random item from the vocabulary. This replacement method yields a new dataset of outfit pairs, where each outfit is labelled as either compatible or non-compatible. The task constitutes a binary classification problem, where we use the area under the curve (AUC) of the receiver operating characteristic (ROC) as the CP metric. To calculate the classification score for BERT, GPT, and the RNN, we compute an outfit probability as $\exp\left(-\operatorname{CE}\left(x;\Theta\right)\right)$ and treat it as a classification score for computing AUC.

\subsubsection{Results}

In Table~\ref{table:results_non_personalized_zalando} we present the results for the different non-personalized algorithms using the Zalando-outfits dataset. The Siamese Nets serve as a baseline and we can see that they consistently perform worse than the rest of the models on all metrics (we do not report on perplexity since they are not a language model). We attribute the worse performance to the following two reasons: first, unlike the rest of the models, on prediction Siamese Nets do not produce probability distribution over the entire vocabulary, but rather each item in the vocabulary must be ranked in isolation. Therefore, the scores the model outputs cannot be directly compared to each other and often sub-optimal choices are made by picking the item with highest score. Second, the generation of negative examples needed by the Siamese Nets is also suboptimal since it relies on the strong assumption that randomly changing items in the outfit always results in set of items that are not compatible.
% We see two reasons for its inferiority compared to the other models: First, it only \textit{looks at} pairs of articles at once. There is no way for it to rate a triple or an $n$-tuple of outfit articles as incompatible, where all pairwise combinations might be compatible. Second, the compatibility of certain commodity groups might be less important than the compatibility of others: Suppose a basic shirt is compatible with all other articles in an outfit, but the combination of shoes and jeans ruins the outfit as a whole, there would be no concept for the Siamese model to assign a high weight to the shoe-pants-compatibility score.

\begin{table}
    \caption{Comparison of non-personalized models on the Zalando outfit dataset.}
    \label{table:results_non_personalized_zalando}
    \begin{tabular}{l r r r r r r}
        \toprule
        Model & PP & CP & FITB@r1 & FITB@r5 & FITB@r25 & FITB@r250 \\
        \midrule
        Siamese Nets & -           & 73.7\% & 0.4\% & 1.3\% & 5.2\% & 23.7\% \\
        LSTM    & 34,290      & 68.6\% & 2.4\% & 5.8\% & 7.9\% & 13.1\% \\ 
        GPT     & \textbf{92} & 96.9\% & 17.7\% & 26.9\% & 37.0\% & 52.2\% \\ 
        BERT     & 182,586 & \textbf{97.9\%} & \textbf{49.3\%} & \textbf{71.7\%} & \textbf{88.2\%} & \textbf{98.6\%} \\ 
        \bottomrule
    \end{tabular}
\end{table}

% !!! TODO GPT and RNN are similar in nature ... wait for RNN results here or discard this paragraph.

BERT and GPT show opposite performance on different metrics. While BERT achieves higher accuracy on the FITB metrics, GPT has a much lower (better) perplexity. That can be explained by the fact that BERT is trained on a task similar to the FITB metric, giving the model a significant advantage, while the GPT is trained on a loss that resembles perplexity, hence performing much better on this metric. Although we expected similar performance between GPT and LSTM, we observed consistently worse performance of the LSTM on both of our outfit datasets.

% On the other hand, BERT never encounters incomplete outfits during training. Hence, the sequential, left-to-right generation perplexity evaluation is presenting it with outfits it has not seen during training, leading to a strong degradation in performance.
% GPT in turn is trained in a sequential manner and has consequently a low perplexity.

On the compatibility task (CP), BERT obtains the best results with 97.9\%, followed very closely by GPT, with 96.9\%. The high AUCs are remarkable given that the training dataset does not contain any negative samples and the models were never explicitly trained on the task of distinguishing compatible and non-compatible outfits.

Our results confirm that, similarly to the NLP domain, GPT is much better suited for generation purposes than BERT. On the other hand, BERT excels at completing outfits with a single missing item. Investigating the usefulness of the contextualized internal representations that BERT computes for each item is a topic left as future work.

In Table~\ref{table:results_non_personalized_zalon} we present the same results on the Zalon outfit dataset. Comparing the results from the two datasets against each other, we see that while they are consistent with respect to the model performance, they are systematically better on the Zalando outfit dataset. While this deserves further investigation, we believe that the difference may be caused by the different item distribution in influencer and stylist outfits. More specifically, the Zalon stylist outfits are created for a personalized customer and a variety of occasions, which means that they are more diverse. This more heterogeneous distribution might be harder for the models to learn. Furthermore, the Zalon outfits in average contain more items than the Zalando outfits.

\begin{table}
    \caption{Comparison of non-personalized models on the Zalon outfit dataset.}
    \label{table:results_non_personalized_zalon}
    \begin{tabular}{l r r r r r r}
        \toprule
        Model & PP & CP & FITB@r1 & FITB@r5 & FITB@r25 & FITB@r250 \\
        \midrule
        Siamese Nets & - & 71.9\% & 0.1\% & 0.2\% & 0.6\% & 4.5\% \\
        LSTM & 28,637 & 64.1\% & 0.7\% & 1.6\% & 2.9\% & 6.8\% \\
        GPT & \textbf{1,212} & \textbf{92.1\%} & 2.4\% & 6.7\% & 15.3\% & 40.8\% \\
        BERT & 9,934 & 89.0\% & \textbf{4.8\%} & \textbf{12.5\%} & \textbf{26.1\%} & \textbf{61.9\%} \\
        \bottomrule
    \end{tabular}
\end{table}

\subsection{Personalized Outfit Generation}

In this section we present and interpret the experimental results of the personalized outfit generation algorithms. We use two different representations of the user context: action sequences (customer click dataset) and a questionnaire answers. We define metrics suitable for outfit recommendation that capture different aspects of the quality of the generated outfits. Finally, we evaluate the algorithms on past click and purchase (kept item) data.

\subsubsection{Implementation Details}

%Each item is represented by the same features as in Section~\ref{sect:experiments_item_compatibility}. 

\begin{itemize}

\item{\textbf{Contextual GPT} and \textbf{Contextual BERT}}: We use the same architecture as in the non-personalized experiments with the addition of the questionnaire embeddings described in Section~\ref{sect:contextual_bert}. Moreover, in order to compare GPT and BERT on our metrics their sampling methods have to be aligned. Since BERT is sampled with the fixed length of the original outfit, we apply the same procedure for GPT. That means, instead of sampling until the stop token is reached, we sample GPT with the fixed length of the original outfit as well.

\item{\textbf{Transformer}}: Both the encoder and the decoder consist of two layers, with 12 attention heads each. The model dimensionality $d_\text{model}$ was selected to be equal to the total size of the input embeddings, namely 216. Each position in the encoder is represented by learned embeddings of the item attributes introduced in the beginning of the section, concatenated with a one-hot representation of the event type (item click, wishlist-change, cart-change) and a normalized scalar value for the action age, counted in number of days between the outfit and the item click. We use dropout of 0.1 and batch size of 64.

\item{\textbf{Personalized Siamese Nets}}: We adapt Siamese Nets introduced in Section~\ref{sect:siamese_nets} to include personalization as follows. For each relevant item in the assortment, we precompute up to 100 outfits and calculate the nearest neighbors between the browsing history of the user and each of the precomputed outfits that contain a particular item the user is currently interacting with. The nearest neighbors are calculated based on Equation~\ref{eq:nearest_neighbors} by using cosine similarity between the image embeddings. We show the top-1 outfit with highest similarity to the customer.

\item{\textbf{S2S LSTM}}: For the encoder and the decoder of the sequence-to-sequence LSTM we use the same setting from the previous section. We sort the outfits by fashion category and train a forward and a backward model in order to be able to generate full outfits from tip to toe.

\end{itemize}

\begin{figure}
\caption{Algorithmic (right) and stylist (left) outfits side-by-side on the product detail page. The item highlighted in red is the anchor item based on which the Transformer model generated the outfit taking the customer's click history into consideration.}
\label{fig:pdp}
\centering
\includegraphics[width=0.7\textwidth]{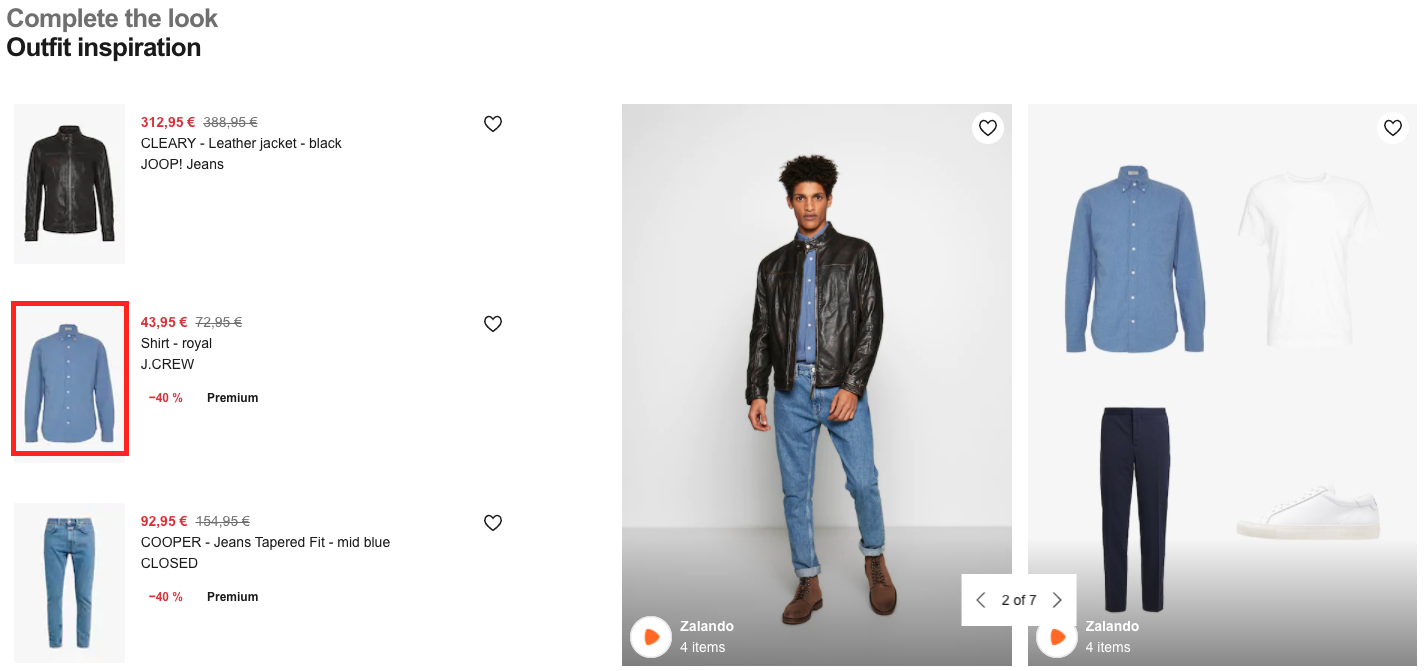}
\end{figure}

We evaluate the action sequence based algorithms on the Zalando click dataset generated from interactions with outfits that complement the main item, which we call anchor, on the product detail page (see Figure \ref{fig:pdp}). We evaluate the questionnaire-based algorithms on Zalon's order and kept items dataset. We use 10\% of 30 days of aggregated action data for evaluation. We use a time-based split, leaving out the last few days of data for evaluation and the rest for training. 

\subsubsection{Metrics}

We use the following set of metrics to asses how diverse the generated outfits are and how well they match individual customer preferences, which is reflected by what the customer has clicked on or purchased. 
  
\paragraph{Fashion attribute click-through rate (CTR)} Many combinations of items could be compatible with a given anchor item since compatibility is defined by multiple factors such as brand, style, color, etc. If an algorithm does not reproduce an exact match, it might be due to the large combination of possible compatible items. Another item might fit perfectly yet be very different visually to what the user has interacted with in the historical click data. To this end, we use proxy metrics for assessing the matches that are based on attributes, in particular the following combinations \emph{brand-category}, \emph{color-category}, \emph{brand-color-category}. For example, if the generated outfit contains an item with the same brand and category as the clicked item, then we consider this a brand-category match. The brand-category hit rate is then the fraction of generated outfits with both, brand and category match.

\paragraph{Fashion attribute keep rate (KR)} Keep rate refers to the fraction of items a user has bought from a shipped full outfit. The proxy metrics for exact KR are defined in the same
way as those for CTR, based on matching fashion attributes. These metrics are used in the experiments on the the Zalon dataset.

\paragraph{Personalization rate} A proxy metric to estimate the ability of the algorithm to personalize, i.e., generate different outfits for different users. It is defined as the ratio of distinct outfits $o$ among the outfits recommended to $n$ different users. It would be 100\% if every user was served a unique outfit.
% Let $S_o$ be the set of outfits recommended to $n$ different users. The personalization rate is defined as $\mathbb{E}\left[\frac{\lvert S_o\rvert}{n}\right]$. 

\paragraph{Item diversity} It is a desirable property of an algorithm to generate outfits with diverse items. The reason for this is: first, showing outfits with a narrow set of items might hurt customer experience, for example, due to obvious repetition of popular items. Second, the outfits should inspire the customer with a broader assortment of items available at the fashion retailer's catalog. We therefore define item diversity as the ratio between the number of unique and the number of total items used to generate all outfits for all users during the offline evaluation.
% TODO for a customer, in total? add a bit more detail - I hope it's clear now.

\subsubsection{Results}

Table~\ref{table:results_personalized_zalando} reports the results of the action-sequence-based algorithms evaluated on the Zalando-click dataset. We use the Siamese Nets algorithm as a baseline since it is widely used in the literature.
%due to its relatively straightforward batch-mode practical implementation.
The Transformer outperforms S2S LSTM and Siamese Nets on all CTR-based metrics. We attribute this to the ability of the model to effectively learn the underlying outfit probability distribution and in the same time learn complex interactions between user click behavior and an outfit the user might be interested in. Moreover, the Transformer and S2S LSTM generate more diverse outfits unlike the Siamese Nets which tends to favor certain items. Finally, the S2S LSTM displays a higher personalization rate albeit significantly lower CTR rate than the Transformer, the main metric for which we optimize. We hypothesize this is due to higher instability of the LSTM in learning the underlying outfit distribution since the LSTM also tends to generate non-valid outfits more often than the Transformer\footnote{A random algorithm would result in close to 100\% personalization rate.}. We leave this further investigation and fine-tuning for future work. 

%This could probably be avoided by more careful hyperparameter tuning which we leave as future work.

% For example, the most common item appeared in more than 7\% of the outfits generated by the Siamese Nets. In the Transformer and S2S LSTM case this figure was 0.8\% and 2.9\% respectively. 

% Marjan: I commented out this text now since we also have LSTM which has even better personalization rate than transformer
% Moreover, we also see that the Transformer performs more than two times better with regards to personalization. The main reason for this is that Transformer builds outfits with taking into account user interactions, while we need to add it on top of the generated outfits with Siamese Networks. In Fig. xxx we can see an example of a Transformer generated outfit.

\begin{table}
    \caption{Personalized action sequence based algorithms evaluated on the Zalando click dataset.}
    \label{table:results_personalized_zalando}
    \begin{tabular}{l r r r}
        \toprule
        Metric & Siamese Nets    & Transformer & Seq-to-Seq LSTM \\
        \midrule
        Brand-category CTR       & 5.8\%       & \textbf{40.8\%} & 9.4\% \\
        Color-category CTR       & 9.3\%       & \textbf{40.2\%} & 12.8\% \\ 
        Brand-color-category CTR & 2.7\%       & \textbf{35.6\%} & 7.4\% \\ 
        Personalizaton rate      & 10.7\%       & 24.1\% & \textbf{51.9\%} \\ 
        Item diversity rate      & 7.7\%       & 31.4\% & \textbf{35.7\%} \\ 
        \bottomrule
    \end{tabular}
\end{table}

In Table~\ref{table:results_personalized_vs_non_personalized_zalon} we compare Contextual GPT and Contextual BERT against their non-personalized counterparts. Here we use the non-personalized metrics from Section~\ref{sect:experiments_item_compatibility} except for the compatibility, which we have excluded since personal context should not significantly affect the ability to distinguish compatible and non-compatible outfits. Regarding the FITB task, we see a significant performance increase for both algorithms: more than 25\% for GPT and more than 10\% for BERT. This shows that the models makes use of the additional information such as preferred color or brand, to predict an item similar to the stylist's choice, who have incorporated this information into their decision process. Furthermore, the perplexity of Contextual GPT  decreased by 40\% for the same reasons, however, BERT's perplexity increased. This can be explained as before by BERT being trained on the FITB task. Namely, the better it gets on the FITB task, the worse its model perplexity gets.

\begin{table}
    \caption{Personalized models compared to their non-personalized counterparts on the Zalon dataset.}
    \label{table:results_personalized_vs_non_personalized_zalon}
    \begin{tabular}{l r r r r r r}
        \toprule
        Model & PP & FITB@1 & FITB@5 & FITB@25 & FITB@250 \\
        \midrule
        GPT & 1,212 & 2.4\% & 6.7\% & 15.3\% & 40.8\% \\
        Contextual GPT & \textbf{728} & 3.1\% & 8.5\% & 19.8\% & 49.5\% \\
        BERT & 9,935 & 4.9\% & 12.5\% & 26.1\% & 61.9\% \\
        Contextual BERT & 15,779 & \textbf{5.9}\% & \textbf{14.5}\% & \textbf{30.9}\% & \textbf{68.1}\% \\
        \bottomrule
    \end{tabular}
\end{table}

In Table~\ref{table:result_personalized_zalando} we compare Contextual GPT and BERT against each other on the more fine-grained personalization metrics defined above. Here we see that GPT performs in general better, i.e., picks items that are more similar to the items the customer has actually kept. While both models benefit from personalization in terms of FITB, this improvement does not seem to translate proportionally in both models in terms of quality of the generated outfits. This might be caused by the different outfit sampling methods: autoregressive generation is a seemingly more efficient and effective generation method than Gibbs sampling that we employ for BERT. We hypothesize that might change if the number of Gibbs sampling iterations is high enough which we plan to investigate in the future.

% This superiority might be explained by the difference in sampling, i.e., autoregressive vs. Gibbs sampling.

\begin{table}
    \caption{Results of the personalized, questionnaire-based algorithms on the Zalon dataset. Metrics are related to purchases; KR stands for keep rate. For example a brand-category KR of 100\% would mean that for every item that was kept by a user there is one item with the same brand and category in the personalized, predicted outfit for that very user.}
    \label{table:result_personalized_zalando}
    \begin{tabular}{l r r}
        \toprule
        Metric & Contextual GPT & Contextual BERT \\
        \midrule
        Brand-category KR       & \textbf{2.0}\% & 0.6\% \\
        Color-category KR       & \textbf{2.3}\% & 1.6\% \\
        Brand-color-category KR & \textbf{0.7}\% & 0.2\% \\
        Personalization rate    & \textbf{0.5}\% & \textbf{0.5}\% \\
        Item diversity rate     & 5.6\%          & \textbf{33.6}\% \\
        \bottomrule
    \end{tabular}
\end{table}

\begin{figure}
\centering
\begin{subfigure}{.5\textwidth}
  \centering
  \includegraphics[width=.95\linewidth]{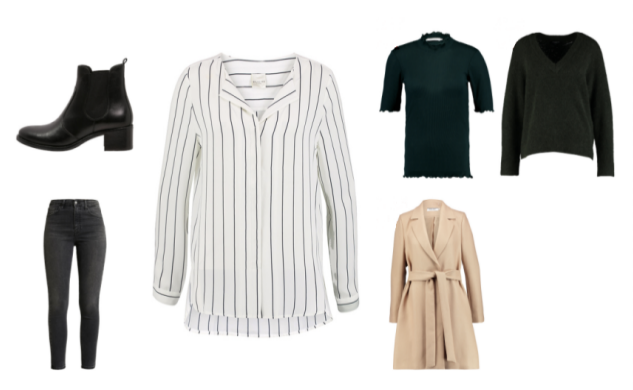}
  \label{fig:sub1}
\end{subfigure}%
\begin{subfigure}{.5\textwidth}
  \centering
  \includegraphics[width=.95\linewidth]{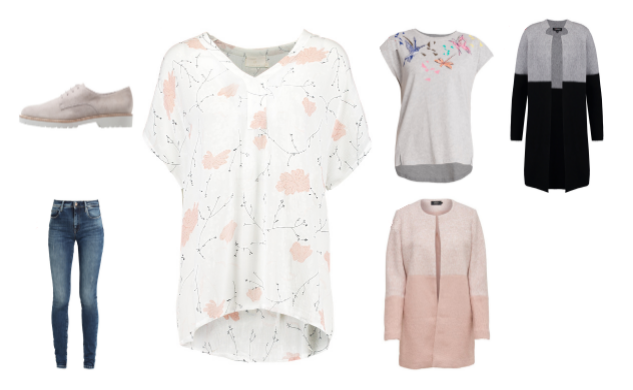}
  \label{fig:sub2}
\end{subfigure}
\caption{Two personalized outfits generated by GPT.}
\label{fig:gpt_outits}
\end{figure}

In Figure \ref{fig:gpt_outits} we show two personalized examples generated by GPT. According to our fashion experts the first one fits perfectly and could have been created by a real stylist. The second one is acceptable, however, the color match between the cardigan and the coat could be improved.

\section{Conclusions and Future work}
\label{sect:conclusions}

In this paper we have provided an experimental evaluation of Siamese Nets, Transformer, GPT, BERT, LSTM, and Seq-to-Seq LSTM on the outfit generation task using real customer data, both for personalized and non-personalized use-case. We have presented new adaptations on BERT, GPT, Siamese Nets and Seq-to-Seq LSTMs for this task and investigated how those have improved the model performance.

Within our extensive experimental results, we have confirmed that GPT outperforms BERT on outfit generation, while showing that BERT provides better performance on the FITB task. Moreover, we have compared personalized and non-personalized approaches, where we have showed that adding personalization does improve the performance of the algorithms with respect to expected customer engagement (e.g., CTR), which confirms that customers are not only looking for compatible outfits, but also for outfits that are of their taste. We have presented that the Transformer outperforms other models in terms of CTR, whereas Seq-to-Seq LSTMs provide higher personalization rate. We have also shown that Siamese Networks are outperformed in both the personalized and non-personalized approaches.

As future work, we plan to investigate more sophisticated methods for personalizing BERT and GPT, such as allowing the models to attend to the personalization context with weights that differ from the self-attention weights instead of prepending it to the input sequence. Such changes have potential to lead to even better personalization. Moreover, we plan to extend our experimental results, while further improving them on the outfit generation task and providing A/B-test results for both GPT and the Transformer.

%Instead of prepending a personalization vector to the input sequence it seems worth investigating to allow the model to attend to the personalization vector with weights that differ from the self-attention weights. We plan to extend our experimental results by a more extensive hyperparameter tuning. Moreover, we plan to provide A/B-test results for both GPT and Transformer, while further improving them on the outfit generation task. 

% also interesting: different representations of the questionnaire. currently answers are formed in subgroups and their embeddings are summed in these sub groups. If one employs a separate attention one could represent each answer separately. 

%% Bibliography
\bibliographystyle{ACM-Reference-Format}
\bibliography{bib_recsys}

\end{document}